\documentclass[conference]{IEEEtran}
\IEEEoverridecommandlockouts
\usepackage{cite}
\usepackage{amsmath,amssymb,amsfonts}
\usepackage{algorithmic}
\usepackage{graphicx}
\usepackage{textcomp}
\usepackage{xcolor}
\def\BibTeX{{\rm B\kern-.05em{\sc i\kern-.025em b}\kern-.08em
    T\kern-.1667em\lower.7ex\hbox{E}\kern-.125emX}}

\usepackage{booktabs}
\usepackage{tabularx}
\usepackage{array}
\usepackage[table]{xcolor}
\usepackage{multirow}
\usepackage{url}
\usepackage{xurl}
\begin{document}

\title{EnvTriCascade: An Environment-Aware Tri-Stage Cascaded Framework for ESDD2 2026 Challenge}

\author{
\IEEEauthorblockN{
Hengyan Huang$^{a*}$, Xiaoxuan Guo$^{a,b*}$, Jiayi Zhou$^b$, Yuankun Xie$^{a,b}$, Jian Liu$^b$, \\
Haonan Cheng$^{a\dag}$, Long Ye$^c$, Qin Zhang$^c$
}

\IEEEauthorblockA{$^a$ \textit{State Key Lab. of Media Convergence and Communication, Communication University of China, Beijing, China}}
\IEEEauthorblockA{$^b$ \textit{Machine Intelligence, Ant Group, Shanghai, China}}
\IEEEauthorblockA{$^c$ \textit{Key Lab. of Media Audio \& Video, Ministry of Education, Communication University of China, Beijing, China}}

\IEEEauthorblockA{
maydayfx@cuc.edu.cn, xiaoxuanguo@mails.cuc.edu.cn, zjy326112@antgroup.com, \\
xieyuankun@cuc.edu.cn, rex.lj@antgroup.com, \{haonancheng, yelong, zhangqin\}@cuc.edu.cn
}

\thanks{$^*$ Equal contribution}
\thanks{$^\dag$ Corresponding author}
}

\maketitle

\begin{abstract}
ADD in real-world scenarios has evolved from speech-only spoofing to more challenging component-level settings, where speech and environmental sounds may be independently manipulated. To tackle this, we propose EnvTriCascade, an Environment-Aware Tri-Stage Cascaded framework for the ESDD2 Challenge. First, a mix-consistency detector provides a binary prior to distinguish original recordings from manipulated mixtures, which calibrates the final decisions. Next, two complementary five-class detectors, leveraging SSLAM+XLS-R and EAT-large+XLS-R representations, extract robust multi-branch features integrated via a cross-branch attention-gated classifier. To enhance robustness against diverse mixing conditions, we incorporate RawBoost augmentation. Trained exclusively on the official CompSpoofV2 dataset, our system achieves a Macro-F1 score of 0.8266 on the test set, significantly outperforming the official baseline and ranking second in the challenge.
\end{abstract}


\section{Introduction}
\label{sec:intro}

Recent advances in text-to-speech~\cite{mai2026magic,ma2026text} and voice conversion~\cite{zheng2026x,yang2026emotion} have greatly increased the realism and accessibility of synthetic audio, raising growing concerns about the authenticity of audio content. Existing audio deepfake detection (ADD) studies have mainly focused on speech anti-spoofing~\cite{huang2025speechfake,huang2025data,guo2026towards}, where systems distinguish bona-fide speech from spoofed speech using spectro-temporal modeling~\cite{jung2022aasist}, self-supervised representations~\cite{zhang24xlsr}, and data augmentation strategies~\cite{tak2022rawboost}. More recently, ADD has been extended beyond speech~\cite{todisco2019asvspoof,guo2026towards,xie2026fake} to broader audio scenarios, including environmental sound deepfake detection~\cite{cheng2024envfake,yin2025esdd,yin2026environmental,yin2025envsdd,guo2026envsslam,chung2026beat2aasist,el2026dfki,cao2026efficient,wei2026domain} and all-type ADD~\cite{xie2025detectall,xie2026add,xie2026interpretable}. This trend reflects a practical challenge: real-world audio is rarely composed of isolated clean speech, but often contains foreground speech, background environmental sounds, and other acoustic events. Therefore, reliable ADD requires not only recognizing spoofing artifacts in speech but also modeling the authenticity of different audio components in complex acoustic scenes.

The ICME 2026 Environment-Aware Speech and Sound Deepfake Detection (ESDD2) challenge further advances this direction by formulating ADD as a component-level classification problem~\cite{zhang2026esdd2}. Unlike conventional whole-audio spoofing detection, ESDD2 considers realistic mixed audio in which speech and environmental sound components may be independently bona-fide or spoofed. The challenge is built upon CompSpoofV2, which extends the component-level audio anti-spoofing setting introduced in CompSpoof~\cite{zhang2026compspoof}. It contains over 250k audio samples with approximately 283 hours of data and defines five target classes: original audio, bona-fide speech with bona-fide environment, spoofed speech with bona-fide environment, bona-fide speech with spoofed environment, and spoofed speech with spoofed environment~\cite{zhang2026esdd2}. The official ranking metric is the overall Macro-F1 score across the five classes. In addition, three Equal Error Rate (EER) metrics are provided for diagnostic analysis, measuring original-vs-mixed discrimination, speech-component spoofing detection, and environmental-component spoofing detection.

In this paper, we present an \textbf{Env}ironment-aware \textbf{Tri}-stage \textbf{Cascade}d Fusion framework, denoted as \textbf{EnvTriCascade}, for ESDD2 Challenge. EnvTriCascade consists of three cascaded inference stages. First, a mix-consistency detector performs an original-vs-mixed binary classification to provide a coarse-grained prior. Second, two complementary five-class detectors conduct component-level classification, where one detector combines SSLAM~\cite{alex2025sslam} with XLS-R~\cite{zhang24xlsr} representations and the other combines EAT-large~\cite{chen2024eat} with XLS-R representations. In each detector, waveform-based and mel-spectrogram-based features are integrated through a cross-branch attention-gated classifier. Finally, the two five-class predictions are fused at the score level and calibrated using the binary prior from the first stage, producing the final decisions.

To further improve robustness under diverse acoustic conditions, we employ RawBoost~\cite{tak2022rawboost} augmentation during training and optimize the overall system with cross-entropy loss using only the official CompSpoofV2 development data, without any external training data. At inference time, the predictions of the two five-class detectors are fused and further calibrated using the mix-consistency output, reducing confusion between original recordings and mixed component-level samples. Our submitted system achieves a Macro-F1 score of 0.8266 on the official test set, significantly outperforming the provided baseline and ranking second in ESDD2 leaderboard.

\begin{figure*}[t]
    \centering
    \includegraphics[width=\linewidth]{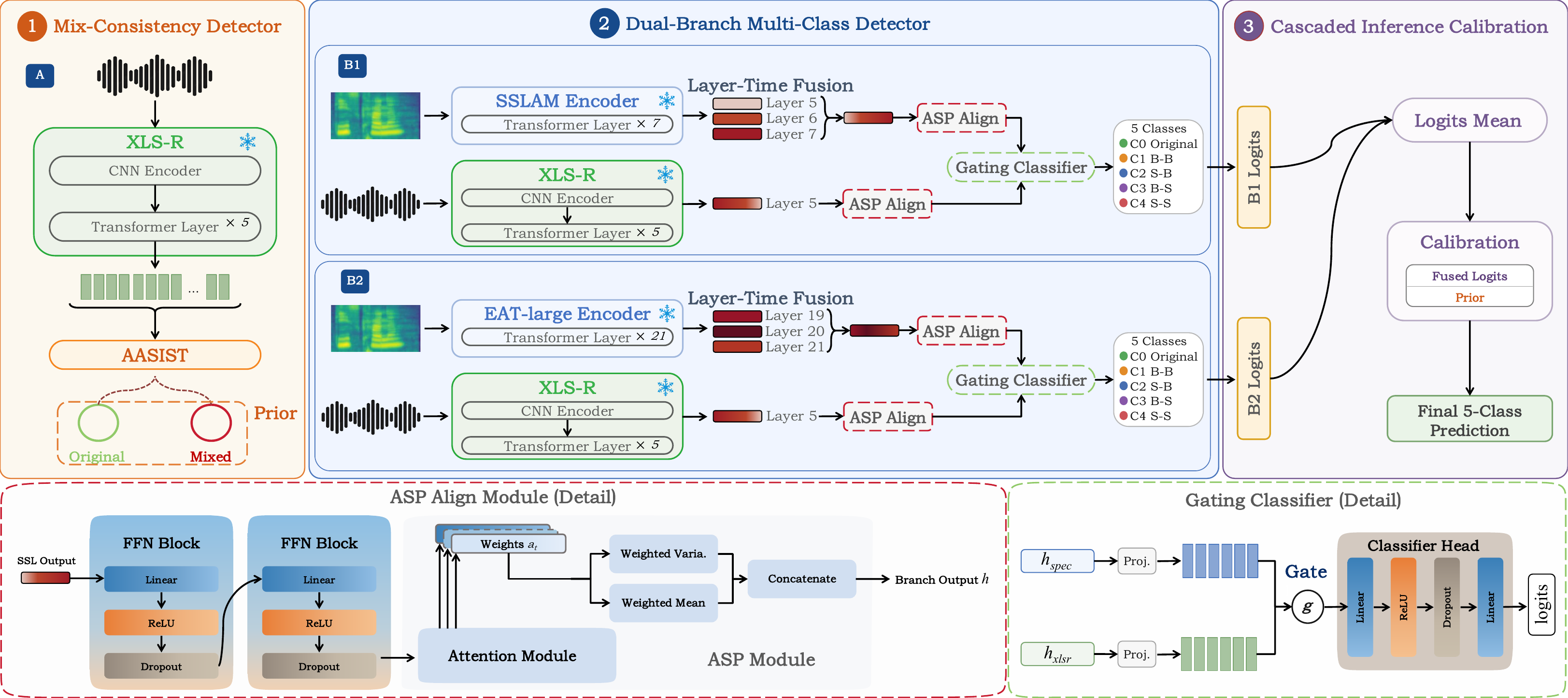}
    \caption{Overall architecture of the proposed EnvTriCascade framework, illustrating the tri-stage inference pipeline: (1) mix-consistency binary screening via System A, (2) dual-branch multi-class ensemble utilizing System B1 and B2, and (3) logic-based decision calibration to yield the final five-class prediction.}
    \label{fig:pipeline}
\end{figure*}

The main contributions of this work are summarized as follows:
\begin{itemize}
    \item We propose EnvTriCascade, an environment-aware ADD framework that achieves the second-place rank in the ICME 2026 ESDD2 Challenge, yielding a Macro-F1 score of 0.8266 on the official test set.

    \item We design a heterogeneous dual-branch multi-class detector to exploit complementary acoustic semantics. Specifically, a layer-time fusion mechanism and a cross-branch gating module are introduced to adaptively extract and integrate spectral and waveform representations.

    \item We develop a tri-stage cascaded inference strategy for component-level spoofing. By employing a mix-consistency detector to provide a binary prior, the framework explicitly bounds and calibrates the multi-class ensemble, effectively mitigating decision conflicts.
\end{itemize}

\section{Methodology}
\label{sec:methodology}
The proposed \textit{EnvTriCascade} framework is designed to disentangle and accurately classify audio clips into five categories corresponding to the authenticity of speech and environmental components: Original (Class 0), Bonafide-Bonafide (Class 1, manipulated but with real components), Spoof-Bonafide (Class 2), Bonafide-Spoof (Class 3), and Spoof-Spoof (Class 4). As illustrated in Fig.~\ref{fig:pipeline}, the system incorporates robust feature extraction, a binary consistency detector, a dual-branch fusion framework, and a decision calibration mechanism.

\subsection{Feature Extractor}
In previous studies on ADD, appropriate representation selection is often more critical than the back-end architecture~\cite{lee2022representation}. To capture both foreground speech and background environmental cues, we employ three distinct self-supervised learning (SSL) backbones. For speech representations, we utilize XLS-R\footnote{\url{https://dl.fbaipublicfiles.com/fairseq/wav2vec/xlsr2_300m.pt}}, which processes raw waveforms and has shown high sensitivity to subtle phonetic spoofing artifacts. Prior work indicates that shallow representations are more effective for speech deepfake detection~\cite{lee2022representation}; therefore, we prune the original 24-layer XLS-R encoder to retain only the first 5 layers. For environmental audio, SSLAM\footnote{\url{https://huggingface.co/ta012/SSLAM_AS2M_Finetuned}} and EAT-large\footnote{\url{https://huggingface.co/worstchan/EAT-large_epoch20_finetune_AS2M}} are adopted. These models employ utterance-frame objectives to effectively capture polyphonic textures and high-level semantics essential for recognizing diverse environmental anomalies. To preserve their generalized representations, all SSL backbones are completely frozen during training.

\subsection{Data Preprocessing}
All audio samples are uniformly resampled to 16 kHz and converted to mono. Temporal alignment is constrained to a 10-second duration, therefore, a repeat-with-jitter strategy is applied to extend shorter inputs. The spectral branch utilizes 128-bin filterbank features with a 10 ms hop size, yielding exactly 1024 frames, which are subsequently normalized.

During the training phase, we inject waveform-level data augmentation using RawBoost, altering signal characteristics to prevent model overfitting to specific acoustic channel conditions. To avoid label noise across branches, the same augmentation seed is synchronized for the raw waveform and the mel-spectrogram inputs within each forward pass.

\begin{figure*}[t]
    \centering
    \begin{minipage}{0.49\textwidth}
        \centering
        \includegraphics[width=\linewidth]{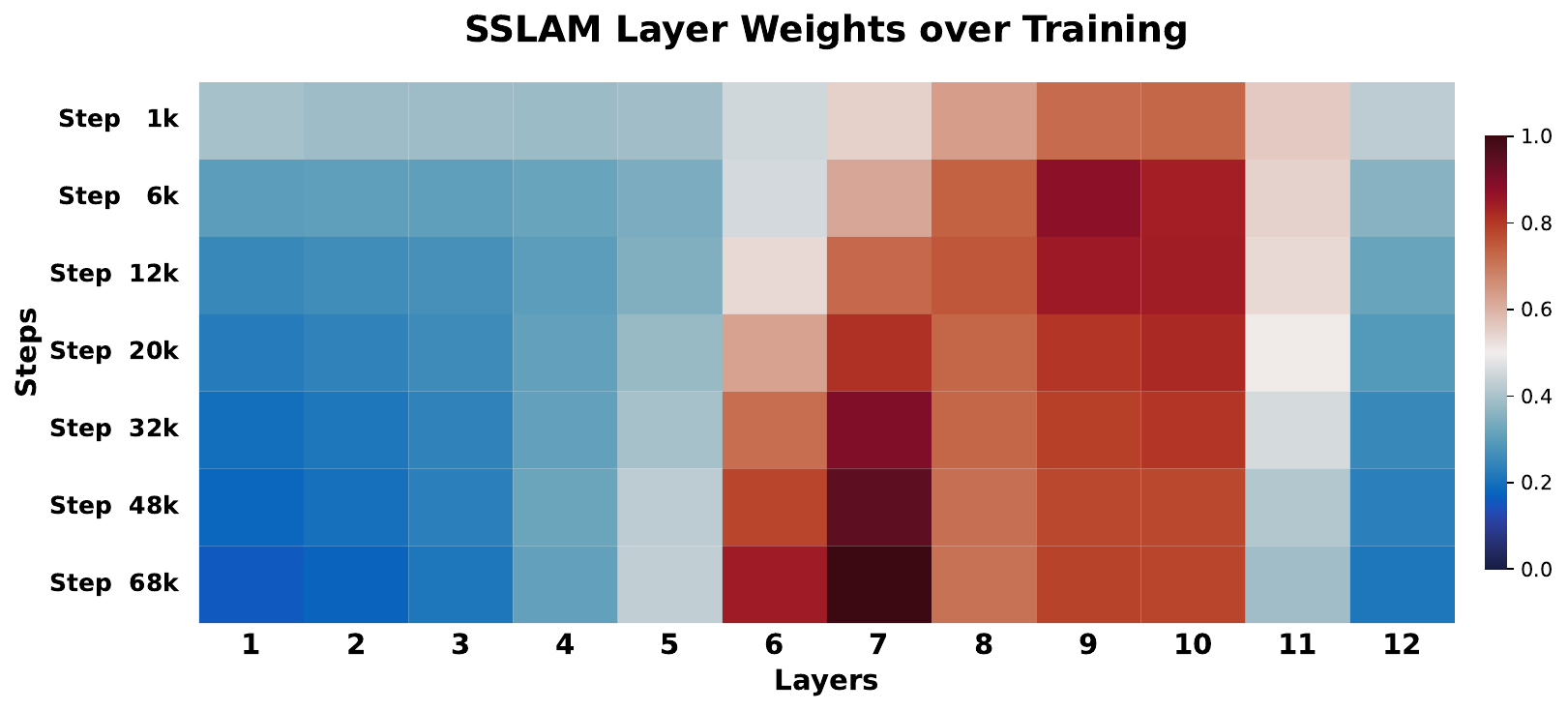}
        \centerline{(a) SSLAM Layer Weights over Training Steps}
    \end{minipage}
    \hfill
    \begin{minipage}{0.49\textwidth}
        \centering
        \includegraphics[width=\linewidth]{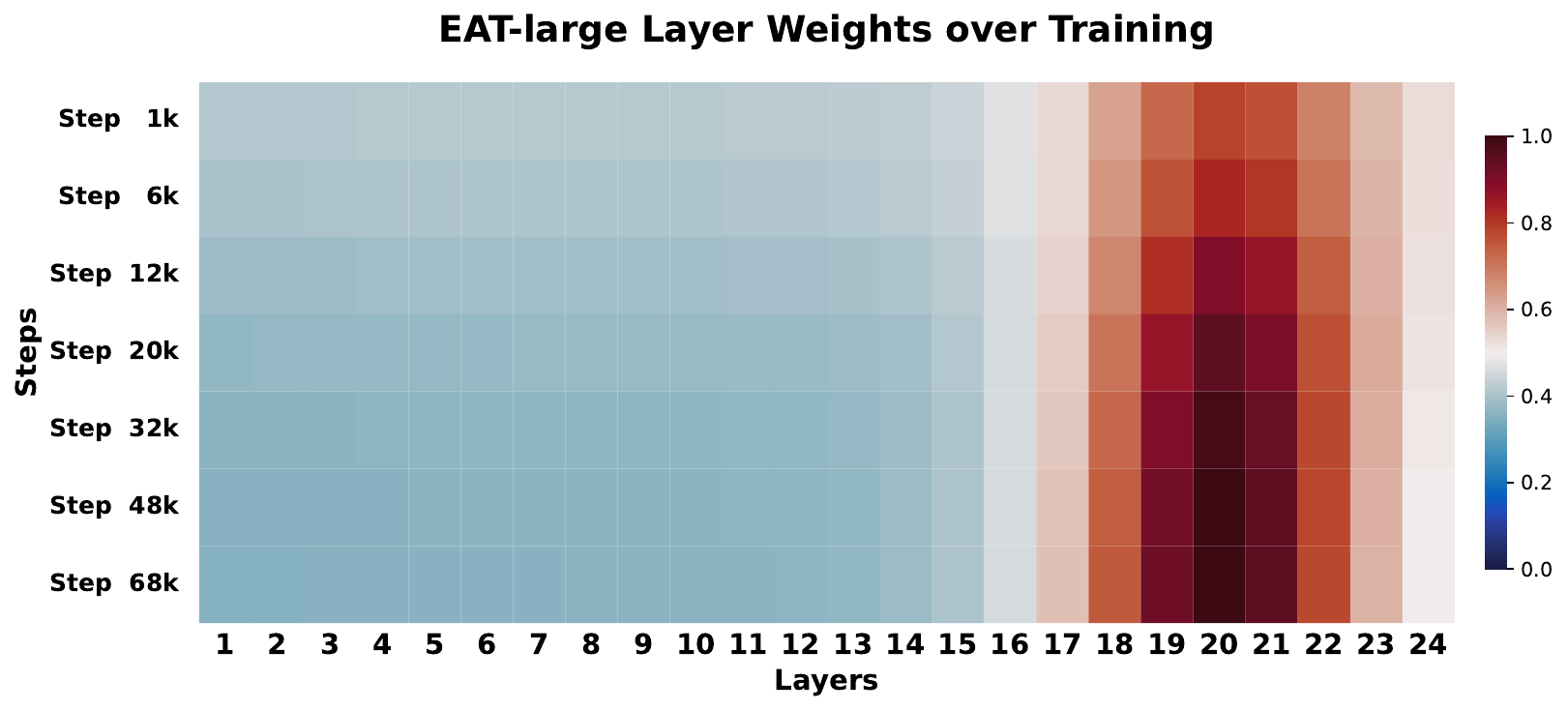}
        \centerline{(b) EAT-large Layer Weights over Training Steps}
    \end{minipage}

    \caption{Evolution of normalized layer-time attention weights during the early training phase. Warmer colors indicate higher contribution scores. The model adaptively concentrates on layers 5--7 for the SSLAM branch (a) and layers 19--21 for the EAT-large branch (b).}
    \label{fig:layer_heatmaps}
\end{figure*}

\subsection{Stage 1: Mix-Consistency Detector}
Empirical analysis of the official ESDD2 baseline revealed that the original class (Class 0) can already be distinguished reliably from manipulated mixtures. Motivated by this observation, we retain binary screening strategy and introduce Stage 1 as a mix-consistency detector (System A). This stage evaluates whether an input is a genuine original recording (Class 0) or a manipulated mixture (Classes 1--4).

For training System A, we use only Class 0 and Class 1 samples. Since Class 0 denotes original recordings and Class 1 denotes mixtures of bona-fide speech and bona-fide environmental sound, this setting encourages the detector to focus on mix-consistency cues rather than synthetic spoofing artifacts. During validation, all five classes are utilized by mapping Class 0 to the original category and Classes 1--4 to the mixed category. 

To strengthen representation learning, we extract features from the 5th layer of frozen XLS-R and process them through AASIST~\cite{jung2022aasist}. This design preserves the effective binary formulation of baseline while improving the feature quality for identifying original audio.





\subsection{Stage 2: Dual-Branch Multi-Class Detector}
For fine-grained multi-class classification, we construct a heterogeneous dual-branch architecture (System B). The spectral branch employs either SSLAM or EAT-large, while the waveform branch utilizes XLS-R.

\subsubsection{Layer-Time Fusion}
Instead of traditional static layer averaging to aggregate features from multiple transformer layers, we propose a layer-time fusion mechanism. Given the multi-layer outputs denoted as $\mathbf{X} \in \mathbb{R}^{L \times T \times D}$, where $L$ is the number of selected layers, $T$ is the sequence length, and $D$ is the feature dimension, we compute a temporal-aware attention score $\alpha_{l,t}$ via a linear projection and a layer-wise softmax:
\begin{equation}
    \alpha_{l,t} = \frac{\exp(\mathbf{W}_{score} \mathbf{X}_{l,t} + b)}{\sum_{j=1}^{L} \exp(\mathbf{W}_{score} \mathbf{X}_{j,t} + b)}
\end{equation}
The fused representation $\mathbf{H} \in \mathbb{R}^{T \times D}$ is then obtained via $\mathbf{H}_{t} = \sum_{l=1}^{L} \alpha_{l,t} \mathbf{X}_{l,t}$. 

To determine the optimal layer subset $L$, we initially conducted a full-layer fusion training phase and monitored the dynamic evolution of the attention weights. Fig.~\ref{fig:layer_heatmaps} illustrates the normalized frame-level attention distributions across layers to quantify their contribution to the final multi-class results. As training advances, the model adaptively concentrates its focus on specific intermediate and deep layers that are most discriminative for deepfake artifacts. For the SSLAM branch (Fig.~\ref{fig:layer_heatmaps}(a)), the attention gradually converges on layers 5, 6, and 7. Conversely, for the EAT-large branch (Fig.~\ref{fig:layer_heatmaps}(b)), the model shifts its focus deeper, peaking at layers 19, 20, and 21. To strike an optimal balance between representation performance and parameter efficiency, we prune the SSL backbones and exclusively retain these highly informative layer subsets for feature extraction.

\subsubsection{Cross-Branch Gating Classifier}
To aggregate the temporal sequence $\mathbf{H}$ into an utterance-level embedding, we employ a Feed-Forward Network (FFN) coupled with an Attentive Statistics Pooling (ASP)~\cite{okabe2018attentive} module. The ASP computes temporal attention weights and outputs the concatenation of the weighted mean and variance, effectively capturing both central tendencies and artifact fluctuations.

The resulting embeddings from the spectral branch $\mathbf{h}_{spec}$ and the waveform branch $\mathbf{h}_{xlsr}$ are linearly projected to a unified dimension $D_f=768$. We then implement an attention gating mechanism to fuse these heterogeneous representations. A learnable sigmoid-activated gate $\mathbf{g}$ performs element-wise weighting:
\begin{equation}
    \mathbf{h}_{fuse} = \mathbf{g} \odot \mathbf{h}_{spec} + (\mathbf{1} - \mathbf{g}) \odot \mathbf{h}_{xlsr}
\end{equation}
where $\odot$ denotes element-wise multiplication. This gating serves as an adaptive selector, allowing the model to dynamically shifts focus between spectral and waveform cues depending on the specific manipulation type. $\mathbf{h}_{fuse}$ is finally passed to the classification head. The architectural details are summarized in Table~\ref{tab:architecture}.

\begin{table*}[htbp]
\centering
\caption{Detailed architecture and tensor transformations of the proposed Dual-Branch Multi-Class Detector (System B).}
\label{tab:architecture}
\setlength{\tabcolsep}{6pt}
\renewcommand{\arraystretch}{1.15}
\small
\begin{minipage}{0.8\textwidth}
\centering
\begin{tabular}{>{\centering\arraybackslash}p{3.2cm}
                >{\centering\arraybackslash}p{6.0cm}
                >{\centering\arraybackslash}p{3.2cm}}
\hline
\textbf{Architecture} & \textbf{Information} & \textbf{Output Shape} \\
\hline
\multicolumn{3}{c}{\textit{Branch 1: SSLAM / EAT (Spectral)}} \\
\hline
SSL Output              & $x_s \in \mathbb{R}^{L_s \times T_s \times D_s}$          & $L_s \times T_s \times D_s$ \\
Layer-Time Fusion  & Layer-wise fusion over $L_s$ layers                            & $T_s \times D_s$ \\
FFN Block          & Linear$(D_s \rightarrow 128)$ + ReLU + Dropout                 & $T_s \times 128$ \\
FFN Block          & Linear$(128 \rightarrow 128)$ + ReLU + Dropout                 & $T_s \times 128$ \\
Attention Block    & Temporal attention over $T_s$ frames                           & $T_s$ \\
ASP                & Mean + variance pooling                                        & $256$ \\
Branch Output      & Spectral branch embedding                                      & $256$ \\
\hline
\multicolumn{3}{c}{\textit{Branch 2: XLS-R (Waveform)}} \\
\hline
SSL Output              & $x_x \in \mathbb{R}^{T_x \times D_x}$                     & $T_x \times D_x$ \\
FFN Block          & Linear$(D_x \rightarrow 128)$ + ReLU + Dropout                 & $T_x \times 128$ \\
FFN Block          & Linear$(128 \rightarrow 128)$ + ReLU + Dropout                 & $T_x \times 128$ \\
Attention Block    & Temporal attention over $T_x$ frames                           & $T_x$ \\
ASP                & Mean + variance pooling                                        & $256$ \\
Branch Output      & Waveform branch embedding                                      & $256$ \\
\hline
\multicolumn{3}{c}{\textit{Fusion \& Classification Head}} \\
\hline
Branch Align       & Linear$(256 \rightarrow D_f)$ per branch                       & $D_f$ \\
\multirow{2}{3.2cm}{\centering Gate}
                   & Linear$(2D_f \rightarrow D_f)$ + ReLU                          &
\multirow{2}{3.2cm}{\centering $D_f$} \\
                   & + Linear$(D_f \rightarrow D_f)$ + Sigmoid                      & \\
Fusion             & $g \odot s + (1-g) \odot x$                                    & $D_f$ \\
\multirow{2}{3.2cm}{\centering Classifier}
                   & Linear$(D_f \rightarrow 128)$ + ReLU + Dropout                 &
\multirow{2}{3.2cm}{\centering $5$} \\
                   & + Linear$(128 \rightarrow 5)$                                  & \\
\hline
\end{tabular}
\end{minipage}
\end{table*}

\begin{table*}[t]
  \centering
  \caption{Performance Comparison of the Proposed Sub-Systems and the Final EnvTriCascade Framework on the ESDD2 Test Set.}
  \label{tab:test_performance}
  \setlength{\tabcolsep}{6pt}
  \renewcommand{\arraystretch}{1.15}
  \small
  \begin{tabular}{>{\centering\arraybackslash}p{7.0cm}
                  >{\centering\arraybackslash}p{2.0cm}
                  >{\centering\arraybackslash}p{2.0cm}}
    \toprule
    \textbf{System} & \textbf{Params (M)} & \textbf{Macro-F1} \\
    \midrule
    Official ESDD2 Baseline & 957.85 & 0.6327 \\
    \midrule
    SSLAM + XLS-R (B1) & 126.52 & 0.7588 \\
    EAT-large + XLS-R (B2) & 337.73 & 0.7544 \\
    Logits Fusion of B1 and B2 (B1+B2) & 464.25 & 0.7707 \\
    \midrule
    Stage-3 calibrated B1 (A+B1) & 203.08 & 0.7966 \\
    Stage-3 calibrated B2 (A+B2) & 414.29 & 0.7944 \\
    \midrule
    \rowcolor{gray!15}
    \textbf{EnvTriCascade (A+B1+B2)} & 540.81 & \textbf{0.8266} \\
    \bottomrule
  \end{tabular}
\end{table*}

\subsection{Stage 3: Tri-Stage Cascaded Inference Calibration}
To mitigate the inherent difficulty of fine-grained classification and maximize detection accuracy, final inference phase orchestrates the aforementioned modules:
\begin{itemize}
    \item \textbf{Step 1 (Prior):} System A evaluates whether the audio is Original (Class 0) or mixed. 
    \item \textbf{Step 2 (Ensemble):} Two variants of the dual-branch system (B1 utilizing SSLAM, and B2 utilizing EAT-large) independently predict the five-class probabilities. Their logits are ensembled via a mean strategy.
    \item \textbf{Step 3 (Calibration):} The system logically bounds the Stage 2 ensemble using the Stage 1 prior. If Stage 1 determines the audio is original, the final prediction is forced to Class 0. Conversely, if Stage 1 detects manipulation but Stage 2 ensemble predicts Class 0, the prediction is overridden by selecting the class with the second-highest logit from Stage 2. This constraint significantly suppresses the false-negative fallback phenomenon.
\end{itemize}

\section{Experiments}
\label{sec:experiments}
\subsection{Dataset and Evaluation Metrics}
We evaluate our method on the curated ESDD2 CompSpoofV2 dataset, comprising 175,361 training samples and 24,864 validation samples. The samples are categorized into the five classes defined in Section \ref{sec:methodology}. 

The primary official evaluation metric is the Macro-F1 score across the five classes, supplemented by EER metrics computed on independent binary attributes, including originality, speech components, and environmental components. It is important to note that while EER is suitable for evaluating single-model probability distributions, the hard logical bounds introduced in our Stage-3 cascade mechanism disrupt the monotonic logits required for standard EER thresholding. Consequently, individual EER scores are not recorded for the final EnvTriCascade system, as our strategy deliberately sacrifices continuous probability distributions to exclusively maximize the exact-match accuracy.

\subsection{Implementation Details}
The proposed system is optimized on a single NVIDIA A100 GPU using the cross-entropy loss function. All models are trained over 50 epochs with a batch size of 32 using the AdamW optimizer. An initial learning rate of $1 \times 10^{-4}$ is utilized, accompanied by a weight decay of $1 \times 10^{-4}$. To ensure training stability and prevent gradient explosion, a linear learning rate warmup is applied over the first 5000 steps, followed by gradient clipping with a maximum norm of 1.0. RawBoost data augmentation is applied dynamically with an activation probability of 50\%, simulating realistic acoustic degradations to enhance robustness.

\subsection{Model Configurations}
Three sub-systems are constructed for the final cascade:
\begin{itemize}
    \item \textbf{System A:} The Stage-1 binary mix-consistency detector is built on pruned XLS-R model with AASIST.
    \item \textbf{System B1:} The Stage-2 dual-branch detector combines a 7-layer SSLAM backbone ($D_s=768$) with XLS-R.
    \item \textbf{System B2:} The Stage-2 dual-branch detector replaces SSLAM with a 21-layer EAT-large backbone ($D_s=1024$) for higher-capacity representations.
\end{itemize}

\subsection{Experimental Results and Analysis}
The experimental results on the official test set are detailed in Table~\ref{tab:test_performance}. The official baseline, despite its massive parameter count, only achieves a Macro-F1 of 0.6327. 

\subsubsection{Effectiveness of Dual-Branch Features}
When evaluated independently, System B1 and System B2 achieved Macro-F1 scores of 0.7588 and 0.7544, respectively. This demonstrates that the layer-time fusion and cross-branch gating mechanism effectively extract discriminative representations from both spectral and waveform domains. Fusing the logits of B1 and B2 (B1+B2) further improves the score to 0.7707, confirming that the different SSLAM and EAT-large backbones provide complementary acoustic representations.

\subsubsection{Impact of Stage-3 Calibration}
A significant performance leap is observed when introducing the mix-consistency detector (System A). Applying the Stage-3 logic calibration to B1 alone (A+B1) yields a substantial absolute improvement of nearly 3.8\% (0.7588 $\rightarrow$ 0.7966), with a similar gain observed for A+B2. These results validate the fine-grained multi-class models often struggle with the ambiguous boundary between original audio and mixtures, a problem that is elegantly mitigated by the binary prior.

\subsubsection{Final System Performance}
By integrating the B1+B2 ensemble with Stage-3 calibration, the proposed \textit{EnvTriCascade} system achieves a Macro-F1 score of \textbf{0.8266}, ranking second place in the challenge. This cascaded design systematically incorporates prior constraints to mitigate false negatives, demonstrating high effectiveness for component-level ADD.

\subsubsection{Error Analysis}
Beyond the Macro-F1 improvements, our empirical observations during the validation phase reveal critical insights into component-level spoofing. We observed that single-branch spectral models frequently confuse Class 0 (Original) with Class 1 (Bonafide-Bonafide mixtures), as both lack synthetic artifacts, differing only in the subtle phase mismatch introduced during the physical mixing process. The Stage-1 mix-consistency detector (System A), utilizing shallow 5th-layer XLS-R representations, proves sensitive to these physical mixing phase discontinuities, successfully separating Class 0 from the rest. 

Furthermore, we found that detecting manipulations in Class 2 (Spoof-Bonafide) is generally more stable than in Class 3 (Bonafide-Spoof). Because background environments typically have lower acoustic energy, their spoofing artifacts are easily overshadowed by genuine speech. To address this challenge, we incorporate the heavier EAT-large branch (System B2) to complement the SSLAM branch. With 21-layer depth and broader receptive field, EAT-large demonstrates a superior capacity to capture low-energy, long-term environmental anomalies, while SSLAM remains effective at modeling polyphonic acoustic cues. Their complementary representations strengthen final score-level ensemble.

\subsubsection{Computational Efficiency}
Although the total parameter count of our final framework reaches 540.81 M, the actual trainable parameter overhead is minimal. By freezing the massive SSL backbones (XLS-R, SSLAM, and EAT-large), the trainable parameters—confined exclusively to the layer-time fusion matrices and the cross-branch gating classifiers—account for about 1.1\% of the total network. This parameter-efficient adaptation prevents catastrophic forgetting of the generalized SSL pre-training and ensures training stability on the CompSpoofV2 dataset.

\section{Conclusion}
\label{sec:conclusion}
In this paper, we introduced an environment-aware tri-stage cascaded framework, EnvTriCascade, for the ESDD2 2026 Challenge. Recognizing the complexity of component-level audio manipulations, we designed a dual-branch detector utilizing dynamic layer-time fusion and cross-branch attention gating to adaptively integrate spectral and waveform representations. Furthermore, we implemented a mix-consistency detector as a binary prior to explicitly calibrate the multi-class ensemble decisions. Experimental evaluations on the CompSpoofV2 dataset demonstrate that our cascaded strategy effectively disentangles heterogeneous artifacts and limits false-negative fallbacks. Ultimately, the proposed system achieves a Macro-F1 score of 0.8266, ranking second in the challenge while remaining highly parameter-efficient, with only 1.1\% of parameters being trainable. Our work provides a practical solution for environment-aware ADD research.

\section*{Acknowledgment}
This work was supported by the Young Elite Scientists Sponsorship Program of the Beijing High Innovation Plan (No. 20250973), the Joint Funds of the Natural Science Foundation of Beijing, China (No. L252143), the Fundamental Research Funds for the Central Universities (No. CUCZD2504), and the National Natural Science Foundation of China (No. 62271455).

\bibliographystyle{IEEEbib}
\bibliography{icme2026references}

\end{document}